# Simultaneous Multiparty Quantum Teleportation Supervised by a Controller

Nikhita Singh[*] and Ravi S. Singh[†]

Photonic Quantum-Information and Quantum Optics Group, Deen Dayal Upadhyaya Gorakhpur University, (Department of Physics), Gorakhpur, (Uttar Pradesh), 273009, India.

e-mail: * singhnikhitabhi6@gmail.com; † yesora27@gmail.com

Corresponding authors: Nikhita Singh and Ravi S. Singh

**Abstract**

Quantum network harbours a technology of multiparty transmission and computation of quantum information. We, here, design a quantum circuit comprising of Hadamard and controlled-Not gates for preparation of a cluster state of symmetric and antisymmetric Bell-pairs involving seventeen-qubits, which is, later on, utilized as a quantum channel. This seventeen-qubits quantum channel is employed to strategize a simultaneous multiparty (quattro-directional) quantum teleportation protocol in which four senders (Alice, Bob, Charlie and David) transmit their arbitrary unknown two-qubit states to respective four receivers (Fancy[1], Fancy[2], Fancy[3] and Fancy[4]) under the supervision of a controller, Elle. After successful accomplishment of the protocol, we assess and compare our scheme with contemporary protocols based on quantum (classical) resource consumption, transmitted qubits, operation complexity and efficiency. We found that our protocol's intrinsic efficiency pegs at 21.65%.

## 1. Introduction

Quantum internet [1], a complex network of quantum networks, is a futuristic emerging technology that promises a paradigmatic shift in the way we share and process quantum information. An intensive foundational theoretical activities and novel applications involving quantum network may be witnessed in burgeoning fields such as distributed quantum computing [2], internet of things [3,4], quantum metrology and quantum sensing [5], quantum cloud computing [6], hybrid quantum network [7], teleportation over fibre network [8], quantum repeaters [9], long distance entanglement purification [10].

Quantum communication, utilizing weird non classical characteristics of quantum systems, spearheads a communication technology in which quantum system is employed as the information-carrier for transmission and surpasses the limits of classical communication technology in terms of engraining information security and increasing information transmission capacity [11]. Quantum communication diverges into fields such as quantum cryptography [12,13], quantum teleportation [14-31], super dense coding [32-34], remote state preparation [35,36], quantum key distribution [37-39], quantum secure direct communication [40-43], quantum state sharing [44,45], quantum-telecloning [46,47], quantum tele-amplification [48,49].

Quantum Teleportation (QT) is the first theoretical protocol, discovered by Bennett et al., in which an arbitrary (unknown) quantum state is transferred from one location to a distant location by a quantum channel, Einstein-Podolsky-Rosen-Bell pair [14], with the assistance of local operation and classical communication. Diversified recent investigations have produced variety of QT-protocols in noisy and noiseless environment, namely, controlled QT [50-54], (symmetric or asymmetric) bi-directional (controlled) QT [55-64], multi-directional QT [65-72]. A port-based QT is also proposed in which the receiver must obtain the transmitted state asymptotically just by picking it at one of its N ports in accordance with the outcome of the sender's square-root measurement at the cost of requiring of N pairs of maximally entangled qubits [73].

Quantum network is a multiparty platform which, on account of no-cloning theorem, demands multidirectional QT [65-72]. In 2016, Li et al. introduced tripartite quantum controlled teleportation protocol in which three senders (Alice, Charlie and Edision) could teleport their three arbitrary single qubits to three different receivers (Bob , David and Ford), simultaneously, by using a single quantum resource under the supervision of Tom. Choudhury and Samanta [66] reported simultaneous perfect teleportation of three two qubit state, i.e., three senders (Alice, Charlie and Edision), each having two qubits information-state, to three different receivers (Bob, David and Ford). In 2020, Zha et al. [68] demonstrated four directional controlled QT via using a single ten-qubit entangled state as a resource in which it is claimed that four senders ( Alice, Bob, Charlie and David ) could teleport their four arbitrary single qubit state to four different receivers (Fancy[1], Fancy[2], Fancy[3], Fancy[4]), respectively, with the assistance of controller, Elle. It is, here, opined that if any one of the parties in aforementioned scheme does not cooperate, four directional controlled QT could not be accomplished. Inquisitively, it may be investigated that in Zha et al.'s scheme [68] Elle cannot make control over all QT process involved therein. Elle can control only QT process between Alice to Fancy[1] and Bob to Fancy[2] but not QT between Charlie to Fancy[3] and David to Fancy[4] [70, vikrram's refs].

In the present work, we propose a scheme for simultaneous multiparty (quattro-directional) controlled QT, involving nine participants four senders (Alice, Bob, Charlie and David) and four receivers (Fancy[1], Fancy[2], Fancy[3], Fancy[4]) and one controller, Elle, in which each sender possesses arbitrary two qubits information-state for transmission and seventeen qubits entangled quantum resource state is utilized to accomplish multiparty (quattro-directional) controlled QT only via Bell state measurement and single-qubit state measurement. Also, we worked out a quantum circuit in which seventeen qubits entangled state can be generated from seventeen single qubit product states by using nine Hadamard and controlled-NOT gates. Furthermore, we made a comparison with contemporary protocols based on resource consumption, transmitted qubits, operation complexity and intrinsic efficiency. It is found that our protocol's intrinsic efficiency pegs at 21.65%.

The remaining part of the paper is structured in following Sections. Section 2.1 is devoted for description of generation of seventeen qubits entangled quantum resource state. Section 2.2 describes the protocol, simultaneous multiparty (quattro-directional) controlled QT. Comparison with contemporary protocols and conclusions are drawn based on efficiency of QT in Section 3.

## 2.1 Preparation of Quantum Resource states-quantum channel

A seventeen-qubit entangled state is employed for implementation of proposed quattro-directional controlled QT scheme as quantum channel. Its generation may be described by following steps:

**1.** We initialized the tensor product of seventeen single qubit state, each qubit is set to be in the state, $|0\rangle$, i.e.,

$$|\psi_0\rangle = |0\rangle_1 \otimes |0\rangle_2 \otimes |0\rangle_3 \otimes |0\rangle_4 \otimes |0\rangle_5 \otimes |0\rangle_6 \otimes |0\rangle_7 \otimes |0\rangle_8 \otimes |0\rangle_9 \otimes |0\rangle_{10} \otimes |0\rangle_{11} \otimes |0\rangle_{12} \otimes |0\rangle_{13} \otimes |0\rangle_{14} \otimes |0\rangle_{15} \otimes |0\rangle_{16} \otimes |0\rangle_{17} \quad (1)$$

**2.** Now, we apply Hadamard gate on qubit 17 to obtain,

$$|\psi_1\rangle = |0\rangle_1 \otimes |0\rangle_2 \otimes |0\rangle_3 \otimes |0\rangle_4 \otimes |0\rangle_5 \otimes |0\rangle_6 \otimes |0\rangle_7 \otimes |0\rangle_8 \otimes |0\rangle_9 \otimes |0\rangle_{10} \otimes |0\rangle_{11} \otimes |0\rangle_{12} \otimes |0\rangle_{13} \otimes |0\rangle_{14} \otimes |0\rangle_{15} \otimes |0\rangle_{16} \otimes \frac{(|0\rangle + |1\rangle)_{17}}{\sqrt{2}} \quad (2)$$

**3.** Next, a CNOT gate is applied with qubit 17 as controlled qubit and rest of qubits 1-16 as target qubit to yield the state-vector of seventeen qubit as

$$|\psi_2\rangle = \frac{1}{\sqrt{2}} \left( |00000000000000000\rangle + |11111111111111111\rangle \right)_{1234567891011121314151617} \quad (3)$$

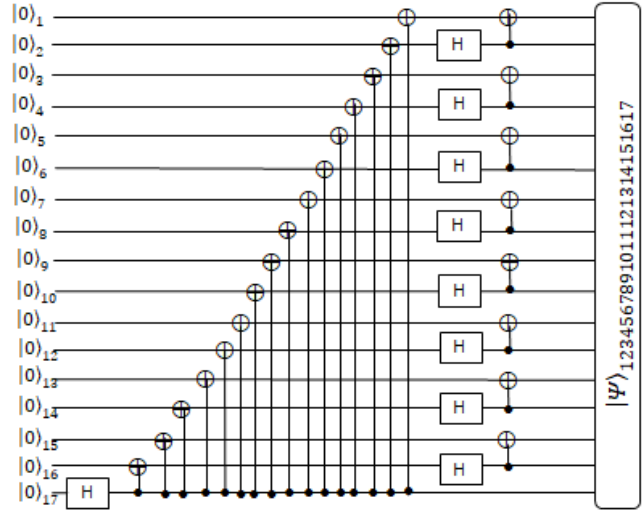

**Fig.1**: Architecture for generation of seventeen-qubits entangled channel employed in proposed simultaneous multiparty controlled QT

**4.** Next, applying Hadamard gates on qubits 2,4,6,8,10,12,14,16 and carrying out CNOT operations on qubits (16,15), (14,13), (12,11), (10,9), (8,7), (6,5), (4,3), (2,1) with first qubit as controlled and second as target qubit, one may produce the required seventeen-qubit entangled state,

$$|\psi\rangle = \frac{1}{\sqrt{2}}[|\kappa^+\rangle_{12}|\kappa^+\rangle_{34}|\kappa^+\rangle_{56}|\kappa^+\rangle_{78}|\kappa^+\rangle_{910}|\kappa^+\rangle_{1112}|\kappa^+\rangle_{1314}|\kappa^+\rangle_{1516}|0\rangle_{17} +$$
$$|\lambda^-\rangle_{12}|\lambda^-\rangle_{34}|\lambda^-\rangle_{56}|\lambda^-\rangle_{78}|\lambda^-\rangle_{910}|\lambda^-\rangle_{1112}|\lambda^-\rangle_{1314}|\lambda^-\rangle_{1516}|1\rangle_{17}] \quad (4)$$

where $|\kappa^+\rangle = \frac{|00\rangle+|11\rangle}{\sqrt{2}}$ and $|\lambda^-\rangle = \frac{|01\rangle-|10\rangle}{\sqrt{2}}$ are Bell-pairs, supplied offline for implementing quattro-directional controlled QT proposed in the present investigation.

## 2.2 Simultaneous multiparty controlled QT

We consider in our proposed scheme for simultaneous multiparty controlled QT, Alice encoded quantum information in arbitrary two qubit (p, q) state,

$$|\phi\rangle_{pq} = (p_{00}|00\rangle + p_{01}|01\rangle + p_{10}|10\rangle + p_{11}|11\rangle)_{pq} \text{ with } \sum_{\mu,\nu=0}^{1}|p_{\mu\nu}|^2 = 1,$$

Bob does the same in arbitrary two qubit (r, s) state,

$$|\phi\rangle_{rs} = (r_{00}|00\rangle + r_{01}|01\rangle + r_{10}|10\rangle + r_{11}|11\rangle)_{rs} \text{ with } \sum_{\mu,\nu=0}^{1}|r_{\mu\nu}|^2 = 1,$$

Charlie's quantum information is possessed by arbitrary two qubit (t, u) state,

$$|\phi\rangle_{tu} = (t_{00}|00\rangle + t_{01}|01\rangle + t_{10}|10\rangle + t_{11}|11\rangle)_{tu} \text{ with } \sum_{\mu,\nu=0}^{1}|t_{\mu\nu}|^2 = 1 \text{ and },$$

And David's quantum information is in an arbitrary two qubit (v,w) state,

$$|\phi\rangle_{vw} = (v_{00}|00\rangle + v_{01}|01\rangle + v_{10}|10\rangle + v_{11}|11\rangle)_{vw}, \text{ with } \sum_{\mu,\nu=0}^{1}|v_{\mu\nu}|^2 = 1.$$

Let us assume that Alice wishes to transmit her quantum state $|\phi\rangle_{pq}$ to Fancy[1], Bobs wishes to transmit his quantum state, $|\phi\rangle_{rs}$ to Fancy[2], Charlie wishes to transmit his quantum state, $|\phi\rangle_{tu}$ to Fancy[3] and David wants to transmit his quantum state $|\phi\rangle_{vw}$ to Fancy[4], simultaneously. In order to establish a quantum link between parties Alice, Bob, Charlie, David, Fancy[1], Fancy[2], Fancy[3] and Fancy[4], we demand that seventeen-qubit entangled state, prepared offline, Eq.(4) is shared as qubits (1, 3), (5, 7), (9, 11), (13, 15) belong to senders, Alice, Bob, Charlie, and David, respectively ; and qubit 17 belong to controller Elle ; and qubits (2, 4), (6, 8), (10, 12), (14, 16) belong to receivers Fancy[1], Fancy[2], Fancy[3] and Fancy[4], respectively. The quantum channel, Eq. (4) may, tacitly, be expressed in more instructive form,

$$|\psi'\rangle = \frac{1}{\sqrt{2}}[|\kappa^+\rangle_{AP}|\kappa^+\rangle_{A'Q}|\kappa^+\rangle_{BR}|\kappa^+\rangle_{B'S}|\kappa^+\rangle_{CT}|\kappa^+\rangle_{C'U}|\kappa^+\rangle_{DV}|\kappa^+\rangle_{D'W}|0\rangle_E +$$
$$|\lambda^-\rangle_{AP}|\lambda^-\rangle_{A'Q}|\lambda^-\rangle_{BR}|\lambda^-\rangle_{B'S}|\lambda^-\rangle_{CT}|\lambda^-\rangle_{C'U}|\lambda^-\rangle_{DV}|\lambda^-\rangle_{D'W}|1\rangle_E] \quad (5)$$

where $(1, 3) = (A, A') \rightarrow$ Alice, $(5, 7) = (B, B') \rightarrow$ Bob, $(9, 11) = (C, C') \rightarrow$ Charlie, $(13, 15) = (D, D') \rightarrow$ David, $(2, 4) = (P, Q) \rightarrow$ Fancy[1], $(6, 8) = (R, S) \rightarrow$ Fancy[2], $(10, 12) = (T, U) \rightarrow$ Fancy[3], $(14, 16) = (V, W) \rightarrow$ Fancy[4] & $17 = (E) \rightarrow$ Elle. Having distributed qubits as described above initial global state-vector of entire system can be expressed as

$$|\chi\rangle_{pqrstuvwAPA'QBRB'SCTC'UDVD'WE}$$
$$= |\phi\rangle_{pq} \otimes |\phi\rangle_{rs} \otimes |\phi\rangle_{tu} \otimes |\phi\rangle_{vw} \otimes |\psi'\rangle_{APA'QBRB'SCTC'UDVD'WE}$$

$$|\chi\rangle_{p-E}$$
$$= (p_{00}|00\rangle + p_{01}|01\rangle + p_{10}|10\rangle + p_{11}|11\rangle)_{pq} \otimes (r_{00}|00\rangle + r_{01}|01\rangle + r_{10}|10\rangle + r_{11}|11\rangle)_{rs}$$
$$\otimes (t_{00}|00\rangle + t_{01}|01\rangle + t_{10}|10\rangle + t_{11}|11\rangle)_{tu} \otimes (v_{00}|00\rangle + v_{01}|01\rangle + v_{10}|10\rangle + v_{11}|11\rangle)_{vw}$$
$$\otimes |\psi'\rangle_{APA'QBRB'SCTC'UDVD'WE}$$

$$|\chi\rangle_{p-E}$$
$$= \frac{1}{\sqrt{2}}\Big[(p_{00}|00\rangle + p_{01}|01\rangle + p_{10}|10\rangle + p_{11}|11\rangle)_{pq} \otimes (r_{00}|00\rangle + r_{01}|01\rangle + r_{10}|10\rangle + r_{11}|11\rangle)_{rs}$$
$$\otimes (t_{00}|00\rangle + t_{01}|01\rangle + t_{10}|10\rangle + t_{11}|11\rangle)_{tu} \otimes (v_{00}|00\rangle + v_{01}|01\rangle + v_{10}|10\rangle + v_{11}|11\rangle)_{vw}$$
$$\otimes |\kappa^+\rangle_{AP}|\kappa^+\rangle_{A'Q}|\kappa^+\rangle_{BR}|\kappa^+\rangle_{B'S}|\kappa^+\rangle_{CT}|\kappa^+\rangle_{C'U}|\kappa^+\rangle_{DV}|\kappa^+\rangle_{D'W}|0\rangle_E \quad (6)$$
$$+ |\phi\rangle_{pq} \otimes |\phi\rangle_{rs} \otimes |\phi\rangle_{tu} \otimes |\phi\rangle_{vw} \otimes |\lambda^-\rangle_{AP}|\lambda^-\rangle_{A'Q}|\lambda^-\rangle_{BR}|\lambda^-\rangle_{B'S}|\lambda^-\rangle_{CT}|\lambda^-\rangle_{C'U}|\lambda^-\rangle_{DV}|\lambda^-\rangle_{D'W}|1\rangle_E\Big]$$

In terms of Bell- pairs, $|\kappa^\pm\rangle$, $|\lambda^\pm\rangle$ and the above equation takes the form, Eq. (A.1) in Appendix.(A), A curious observation of Eq. (A.1) demonstrates that it can be put into a generalized instructive form,

$$|\chi\rangle = \frac{1}{64\sqrt{2}}\Big[\sum_{g,h,i,j,k,l,m,n,=0}^{3}\sum_{z=o}^{1}|\zeta^{(g)}\rangle_{pA}|\zeta^{(h)}\rangle_{qA'}|\zeta^{(i)}\rangle_{rB}|\zeta^{(j)}\rangle_{sB'}|\zeta^{(k)}\rangle_{tC}|\zeta^{(l)}\rangle_{uC'}|\zeta^{(m)}\rangle_{vD}|\zeta^{(n)}\rangle_{wD'}$$
$$|U^{(g,h)}\rangle_{PQ}|U^{(i,j)}\rangle_{RS}|U^{(k,l)}\rangle_{TU}|U^{(m,n)}\rangle_{VW}|\varphi\rangle_{PQ}|\varphi\rangle_{RS}|\varphi\rangle_{TU}|\varphi\rangle_{VW}|z\rangle_E\Big] \quad (7)$$

where, $|\zeta^{(0)}\rangle = |\kappa^+\rangle, |\zeta^{(1)}\rangle = |\kappa^-\rangle, |\zeta^{(2)}\rangle = |\lambda^+\rangle, |\zeta^{(3)}\rangle = |\lambda^-\rangle$ & unitary operations, $U_{PQ}^{(g,h)}$, $U_{RS}^{(i,j)}$, $U_{TU}^{(k,l)}$, $U_{VW}^{(m,n)}$ performed by receivers (Fancy[1], Fancy[2], Fancy[3] & Fancy[4]) are tabulated in Tables 1 and 2 of Appendix(B). Strategy for simultaneous multiparty controlled QT protocol can be described in following steps:

**Step**-1: Alice performs, firstly, Bell-state measurement (BSM) on her qubit pair (p, A) & (q, A′) in the Bell basis, $\{|\kappa^\pm\rangle, |\lambda^\pm\rangle\}$ and communicate her measurement-results to Fancy[1] via classical channel.

**Step**-2: At the same time, Bob, Charlie & David perform BSM on his own qubits-pairs $(r, B)$ & $(s, B')$ , $(t, C)$ & $(u, C')$ and $(v, D)$ & $(w, D')$, respectively, in the Bell basis $\{|\kappa^\pm\rangle, |\lambda^\pm\rangle\}$ and communicate his measurement-results to Fancy[2], Fancy[3] & Fancy[4], respectively, via classical channel.

**Step**-3: Finally, controller, Elle performs single-qubit von-Neuman measurement on his qubit E in the computational basis $\{|0\rangle, |1\rangle\}$ and communicate his measurement result to all the receivers Fancy[1], Fancy[2], Fancy[3] & Fancy[4], respectively, via classical channel.

As an example, Suppose Alice's measurement result is $|\kappa^+\rangle_{pA}, |\kappa^+\rangle_{qA'}$ and, at the same time, if Bob's, Charlie's and David's measurement-results are $|\kappa^+\rangle_{rB}, |\kappa^+\rangle_{sB'}$, $|\kappa^+\rangle_{tC}, |\kappa^+\rangle_{uC'}$ and $|\kappa^+\rangle_{vD}, |\kappa^+\rangle_{wD'}$, respectively, the collapsed state of the remaining qubits ($P, Q, R, S, T, U, V, W, E$) read as

$$|\varphi\rangle_{PQRSTUVWE} = \Big[ \big( p_{00}|00\rangle + p_{01}|01\rangle + p_{10}|10\rangle + p_{11}|11\rangle \big)_{PQ} \otimes \big( r_{00}|00\rangle + r_{01}|01\rangle + r_{10}|10\rangle + r_{11}|11\rangle \big)_{RS}$$
$$\otimes \big( t_{00}|00\rangle + t_{01}|01\rangle + t_{10}|10\rangle + t_{11}|11\rangle \big)_{TU} \otimes \big( v_{00}|00\rangle + v_{01}|01\rangle + v_{10}|10\rangle + v_{11}|11\rangle \big)_{VW} |0\rangle_E \quad (8)$$
$$+ \big( p_{00}|11\rangle - p_{01}|10\rangle - p_{10}|01\rangle + p_{11}|00\rangle \big)_{PQ} \otimes \big( r_{00}|11\rangle - r_{01}|10\rangle - r_{10}|01\rangle + r_{11}|00\rangle \big)_{RS}$$
$$\otimes \big( t_{00}|11\rangle - t_{01}|10\rangle - t_{10}|01\rangle + t_{11}|00\rangle \big)_{TU} \otimes \big( v_{00}|11\rangle - v_{01}|10\rangle - v_{10}|01\rangle + v_{11}|00\rangle \big)_{VW} |1\rangle_E \Big],$$

with probability $1/2^{16}$. From the Eq. (8), it is evident that none of the receivers can reconstruct the desired state without cooperative assistance of the controller, Elle. If the controller, Elle's measurement outcome is $|0\rangle_E$, collapsed state of qubits $(P, Q, R, S, T, U, V, W)$ yielded as

$$|\phi_0\rangle_{PQRSTUV} = \big( p_{00}|00\rangle + p_{01}|01\rangle + p_{10}|10\rangle + p_{11}|11\rangle \big)_{PQ} \otimes \big( r_{00}|00\rangle + r_{01}|01\rangle + r_{10}|10\rangle + r_{11}|11\rangle \big)_{RS}$$
$$\otimes \big( t_{00}|00\rangle + t_{01}|01\rangle + t_{10}|10\rangle + t_{11}|11\rangle \big)_{TU} \otimes \big( v_{00}|00\rangle + v_{01}|01\rangle + v_{10}|10\rangle + v_{11}|11\rangle \big)_{VW} \quad (9)$$

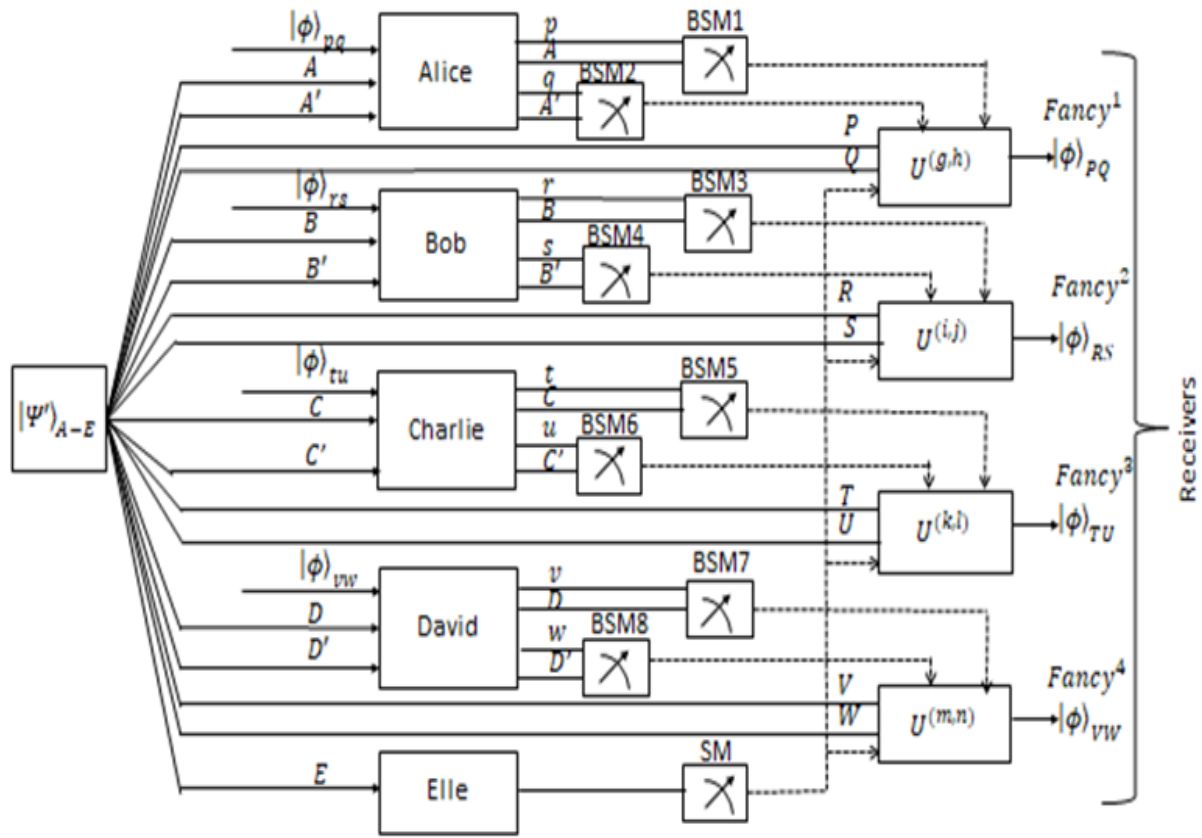

**Fig.2**: Schematic quantum circuit for Simultaneous multiparty QT protocol. Here, $|\psi'\rangle_{A-E} = |\psi'\rangle_{APA'QBRB'SCTC'UDVD'WE} \rightarrow$ Quantum Channel, Information state possess by senders (Alice, Bob, Charlie, David) $|\phi\rangle_{pq}, |\phi\rangle_{rs}, |\phi\rangle_{tu}, |\phi\rangle_{vw}$, respectively, BSM is acronym for Bell-state measurement on two qubits (see text), SM represents single-qubit measurement on qubit, and $U^{(g,h)}, U^{(i,j)}, U^{(k,l)}, U^{(m,n)}$, are unitary operations performed by receivers Fancy[1], Fancy[2], Fancy[3] & Fancy[4]. Dashed lines are for classical-bits communications.

In this case receivers would perform unitary operations $I \otimes I$ on their qubits to reconstruct the original state of senders (Alice, Bob, Charlie and David), and, hence, quattro-directional QT gets accomplished. If the controller, Elle's measurement result is $|1\rangle_E$, the collapsed state of qubits $(P,Q,R,S,T,U,V,W)$ is given by

$$|\phi_0\rangle_{PQRSTUVW} = (p_{00}|11\rangle - p_{01}|10\rangle - p_{10}|01\rangle + p_{11}|00\rangle)_{PQ} \otimes (r_{00}|11\rangle - r_{01}|10\rangle - r_{10}|01\rangle + r_{11}|00\rangle)_{RS} \\ \otimes (t_{00}|11\rangle - t_{01}|10\rangle - t_{10}|01\rangle + t_{11}|00\rangle)_{TU} \otimes (v_{00}|11\rangle - v_{01}|10\rangle - v_{10}|01\rangle + v_{11}|00\rangle)_{VW}. \quad (10)$$

In this case receivers will perform suitable unitary operations, $\sigma_x\sigma_z \otimes \sigma_x\sigma_z$ on their qubits to reconstruct the original state. For other cases, the receivers (Fancy[1], Fancy[2], Fancy[3] & Fancy[4]) will perform appropriate unitary operations $\widetilde{U}_{PQ}^{(g,h)}, \widetilde{U}_{RS}^{(i,j)}, \widetilde{U}_{TU}^{(k,l)}$ & $\widetilde{U}_{VW}^{(m,n)}$, respectively, depending upon variant measurement-results $(g,h,i,j,k,l,m,n)$ received from respective senders (Alice, Bob, Charlie & David), where $\widetilde{U}$ is the Hermitian conjugate operator corresponding to unitary operator $U$.

## 3 Comparison and Conclusions

The intrinsic efficiency is computed to compare our scheme with other protocols. Intrinsic efficiency is defined as $\tau = q_s/(q_u + b_t)$ in which $q_s$ represents the number of quantum information bits transmitted, $q_u$ represents number of qubits present in quantum channel and $b_t$ represents number of classical bits transmitted for completion of protocol. A comparison Table 3 demonstrating intrinsic efficiency, quantum-resource consumption and classical-resource consumption are presented.

Conclusively, in this paper, we have designed a quantum circuit for preparation of seventeen-qubit entangled state employed as the quantum channel which is used to realize simultaneous multiparty controlled quantum teleportation protocol. In our approach, four senders Alice, Bob, Charlie & David may transmit their unknown arbitrary two qubit states simultaneously to four respective receivers Fancy[1], Fancy[2], Fancy[3] and Fancy[4], respectively under the controller, Elle where only experimentally feasible BSMs and single qubit von-Neuman measurements are required. The present investigation may be applied for transmission of quantum information within a quantum network.

Table 3: Comparisons amongst Contemporary multidirectional Protocols.

| Refs. | Participants | | | QIBT | QRC | CRC | Measurements | | Efficiency ($\eta$) |
|---|---|---|---|---|---|---|---|---|---|
| | Sender | receiver | controller | Qubit | Qubit | | BSM | SM | |
| [65] | 3 | 3 | 1 | 3 | 7 | 9 | 3 | 1 | 18.75 |
| [68] | 4 | 4 | 1 | 4 | 10 | 16 | 4 | 2 | 15.38 |
| [70] | 4 | 4 | 1 | 4 | 9 | 21 | 4 | 1 | 19.04 |
| ours | 4 | 4 | 1 | 8 | 17 | 37 | 8 | 1 | 21.65 |

QIBT (Quantum-information bits (to be) transmitted), QRC (Quantum resource consumption), CRC (Classical resource consumption), BSM (Bell State measurement), SM (Single qubit measurement).

## Acknowledgements

We acknowledge Prof. Lev Vaidman, School of Physics and Astronomy, Tel Aviv University, Israel and Prof. Mark M. Wilde, School of Electrical and Computer Engineering Cornell University, New York for drawing our attention toward their investigations on continuous variable quantum teleportation and bidirectional quantum teleportation.

**Appendix.A** The global state vector of entire qubits's system (see the text below Eq.6)

$$
\begin{aligned}
|\chi\rangle_{p-E} = &\Big\{|\kappa^+\rangle_{pA}|\kappa^+\rangle_{qA'}|\eta^{(1)}\rangle_{PQ} + |\kappa^+\rangle_{pA}|\kappa^-\rangle_{qA'}|\eta^{(2)}\rangle_{PQ} + |\kappa^-\rangle_{pA}|\kappa^+\rangle_{qA'}|\eta^{(3)}\rangle_{PQ} + |\kappa^-\rangle_{pA}|\kappa^-\rangle_{qA'}|\eta^{(4)}\rangle_{PQ} \\
&+ |\kappa^+\rangle_{pA}|\lambda^+\rangle_{qA'}|\eta^{(5)}\rangle_{PQ} + |\kappa^+\rangle_{pA}|\lambda^-\rangle_{qA'}|\eta^{(6)}\rangle_{PQ} + |\kappa^-\rangle_{pA}|\lambda^+\rangle_{qA'}|\eta^{(7)}\rangle_{PQ} + |\kappa^-\rangle_{pA}|\lambda^-\rangle_{qA'}|\eta^{(8)}\rangle_{PQ} \\
&+ |\lambda^+\rangle_{pA}|\kappa^+\rangle_{qA'}|\eta^{(9)}\rangle_{PQ} + |\lambda^+\rangle_{pA}|\kappa^-\rangle_{qA'}|\eta^{(10)}\rangle_{PQ} + |\lambda^-\rangle_{pA}|\kappa^+\rangle_{qA'}|\eta^{(11)}\rangle_{PQ} + |\lambda^-\rangle_{pA}|\kappa^-\rangle_{qA'}|\eta^{(12)}\rangle_{PQ} \\
&+ |\lambda^+\rangle_{pA}|\lambda^+\rangle_{qA'}|\eta^{(13)}\rangle_{PQ} + |\lambda^+\rangle_{pA}|\lambda^-\rangle_{qA'}|\eta^{(14)}\rangle_{PQ} + |\lambda^-\rangle_{pA}|\lambda^+\rangle_{qA'}|\eta^{(15)}\rangle_{PQ} + |\lambda^-\rangle_{pA}|\lambda^-\rangle_{qA'}|\eta^{(16)}\rangle_{PQ}\Big\}\otimes \\
&\Big\{|\kappa^+\rangle_{rB}|\kappa^+\rangle_{sB'}|\eta^{(17)}\rangle_{RS} + |\kappa^+\rangle_{rB}|\kappa^-\rangle_{sB'}|\eta^{(18)}\rangle_{RS} + |\kappa^-\rangle_{rB}|\kappa^+\rangle_{sB'}|\eta^{(19)}\rangle_{RS} + |\kappa^-\rangle_{rB}|\kappa^-\rangle_{sB'}|\eta^{(20)}\rangle_{RS} \\
&+ |\kappa^+\rangle_{rB}|\lambda^+\rangle_{sB'}|\eta^{(21)}\rangle_{RS} + |\kappa^+\rangle_{rB}|\lambda^-\rangle_{sB'}|\eta^{(22)}\rangle_{RS} + |\kappa^-\rangle_{rB}|\lambda^+\rangle_{sB'}|\eta^{(23)}\rangle_{RS} + |\kappa^-\rangle_{rB}|\lambda^-\rangle_{sB'}|\eta^{(24)}\rangle_{RS} \\
&+ |\lambda^+\rangle_{rB}|\kappa^+\rangle_{sB'}|\eta^{(25)}\rangle_{RS} + |\lambda^+\rangle_{rB}|\kappa^-\rangle_{sB'}|\eta^{(26)}\rangle_{RS} + |\lambda^-\rangle_{rB}|\kappa^+\rangle_{sB'}|\eta^{(27)}\rangle_{RS} + |\lambda^-\rangle_{rB}|\kappa^-\rangle_{sB'}|\eta^{(28)}\rangle_{RS} \\
&+ |\lambda^+\rangle_{rB}|\lambda^+\rangle_{sB'}|\eta^{(29)}\rangle_{RS} + |\lambda^+\rangle_{rB}|\lambda^-\rangle_{sB'}|\eta^{(30)}\rangle_{RS} + |\lambda^-\rangle_{rB}|\lambda^+\rangle_{sB'}|\eta^{(31)}\rangle_{RS} + |\lambda^-\rangle_{rB}|\lambda^-\rangle_{sB'}|\eta^{(32)}\rangle_{RS}\Big\}\otimes \\
&\Big\{|\kappa^+\rangle_{tC}|\kappa^+\rangle_{uC'}|\eta^{(33)}\rangle_{TU} + |\kappa^+\rangle_{tC}|\kappa^-\rangle_{uC'}|\eta^{(34)}\rangle_{TU} + |\kappa^-\rangle_{tC}|\kappa^+\rangle_{uC'}|\eta^{(35)}\rangle_{TU} + |\kappa^-\rangle_{tC}|\kappa^-\rangle_{uC'}|\eta^{(36)}\rangle_{TU} \\
&+ |\kappa^+\rangle_{tC}|\lambda^+\rangle_{uC'}|\eta^{(37)}\rangle_{TU} + |\kappa^+\rangle_{tC}|\lambda^-\rangle_{uC'}|\eta^{(38)}\rangle_{TU} + |\kappa^-\rangle_{tC}|\lambda^+\rangle_{uC'}|\eta^{(39)}\rangle_{TU} + |\kappa^-\rangle_{tC}|\lambda^-\rangle_{uC'}|\eta^{(40)}\rangle_{TU} \\
&+ |\lambda^+\rangle_{tC}|\kappa^+\rangle_{uC'}|\eta^{(41)}\rangle_{TU} + |\lambda^+\rangle_{tC}|\kappa^-\rangle_{uC'}|\eta^{(42)}\rangle_{TU} + |\lambda^-\rangle_{tC}|\kappa^+\rangle_{uC'}|\eta^{(43)}\rangle_{TU} + |\lambda^-\rangle_{tC}|\kappa^-\rangle_{uC'}|\eta^{(44)}\rangle_{TU} \\
&+ |\lambda^+\rangle_{tC}|\lambda^+\rangle_{uC'}|\eta^{(45)'}\rangle_{TU} + |\lambda^+\rangle_{tC}|\lambda^-\rangle_{uC'}|\eta^{(46)}\rangle_{TU} + |\lambda^-\rangle_{tC}|\lambda^+\rangle_{uC'}|\eta^{(47)}\rangle_{TU} + |\lambda^-\rangle_{tC}|\lambda^-\rangle_{uC'}|\eta^{(48)}\rangle_{TU}\Big\}\otimes \\
&\Big\{|\kappa^+\rangle_{vD}|\kappa^+\rangle_{wD'}|\eta^{(49)}\rangle_{VW} + |\kappa^+\rangle_{vD}|\kappa^-\rangle_{wD'}|\eta^{(50)}\rangle_{VW} + |\kappa^-\rangle_{vD}|\kappa^+\rangle_{wD'}|\eta^{(51)}\rangle_{VW} + |\kappa^-\rangle_{vD}|\kappa^-\rangle_{wD'}|\eta^{(52)}\rangle_{VW} \\
&+ |\kappa^+\rangle_{vD}|\lambda^+\rangle_{wD'}|\eta^{(53)}\rangle_{VW} + |\kappa^+\rangle_{vD}|\lambda^-\rangle_{wD'}|\eta^{(54)}\rangle_{VW} + |\kappa^-\rangle_{vD}|\lambda^+\rangle_{wD'}|\eta^{(55)}\rangle_{VW} + |\kappa^-\rangle_{vD}|\lambda^-\rangle_{wD'}|\eta^{(56)}\rangle_{VW} \\
&+ |\lambda^+\rangle_{vD}|\kappa^+\rangle_{wD'}|\eta^{(57)}\rangle_{VW} + |\lambda^+\rangle_{vD}|\kappa^-\rangle_{wD'}|\eta^{(58)}\rangle_{VW} + |\lambda^-\rangle_{vD}|\kappa^+\rangle_{wD'}|\eta^{(59)}\rangle_{VW} + |\lambda^-\rangle_{vD}|\kappa^-\rangle_{wD'}|\eta^{(60)}\rangle_{VW} \\
&+ |\lambda^+\rangle_{vD}|\lambda^+\rangle_{wD'}|\eta^{(61)}\rangle_{VW} + |\lambda^+\rangle_{vD}|\lambda^-\rangle_{wD'}|\eta^{(62)}\rangle_{VW} + |\lambda^-\rangle_{vD}|\lambda^+\rangle_{wD'}|\eta^{(63)}\rangle_{VW} + |\lambda^-\rangle_{vD}|\lambda^-\rangle_{wD'}|\eta^{(64)}\rangle_{VW}\Big\}|0\rangle_E \\
&+ \Big\{|\kappa^+\rangle_{pA}|\kappa^+\rangle_{qA'}|\eta'^{(1)}\rangle_{PQ} + |\kappa^+\rangle_{pA}|\kappa^-\rangle_{qA'}|\eta'^{(2)}\rangle_{PQ} + |\kappa^-\rangle_{pA}|\kappa^+\rangle_{qA'}|\eta'^{(3)}\rangle_{PQ} + |\kappa^-\rangle_{pA}|\kappa^-\rangle_{qA'}|\eta'^{(4)}\rangle_{PQ} \\
&+ |\kappa^+\rangle_{pA}|\lambda^+\rangle_{qA'}|\eta'^{(5)}\rangle_{PQ} + |\kappa^+\rangle_{pA}|\lambda^-\rangle_{qA'}|\eta'^{(6)}\rangle_{PQ} + |\kappa^-\rangle_{pA}|\lambda^+\rangle_{qA'}|\eta'^{(7)}\rangle_{PQ} + |\kappa^-\rangle_{pA}|\lambda^-\rangle_{qA'}|\eta'^{(8)}\rangle_{PQ} \\
&+ |\lambda^+\rangle_{pA}|\kappa^+\rangle_{qA'}|\eta'^{(9)}\rangle_{PQ} + |\lambda^+\rangle_{pA}|\kappa^-\rangle_{qA'}|\eta'^{(10)}\rangle_{PQ} + |\lambda^-\rangle_{pA}|\kappa^+\rangle_{qA'}|\eta'^{(11)}\rangle_{PQ} + |\lambda^-\rangle_{pA}|\kappa^-\rangle_{qA'}|\eta'^{(12)}\rangle_{PQ} \\
&+ |\lambda^+\rangle_{pA}|\lambda^+\rangle_{qA'}|\eta'^{(13)}\rangle_{PQ} + |\lambda^+\rangle_{pA}|\lambda^-\rangle_{qA'}|\eta'^{(14)}\rangle_{PQ} + |\lambda^-\rangle_{pA}|\lambda^+\rangle_{qA'}|\eta'^{(15)}\rangle_{PQ} + |\lambda^-\rangle_{pA}|\lambda^-\rangle_{qA'}|\eta'^{(16)}\rangle_{PQ}\Big\}\otimes \\
&\Big\{|\kappa^+\rangle_{rB}|\kappa^+\rangle_{sB'}|\eta'^{(17)}\rangle_{RS} + |\kappa^+\rangle_{rB}|\kappa^-\rangle_{sB'}|\eta'^{(18)}\rangle_{RS} + |\kappa^-\rangle_{rB}|\kappa^+\rangle_{sB'}|\eta'^{(19)}\rangle_{RS} + |\kappa^-\rangle_{rB}|\kappa^-\rangle_{sB'}|\eta'^{(20)}\rangle_{RS} \\
&+ |\kappa^+\rangle_{rB}|\lambda^+\rangle_{sB'}|\eta'^{(21)}\rangle_{RS} + |\kappa^+\rangle_{rB}|\lambda^-\rangle_{sB'}|\eta'^{(22)}\rangle_{RS} + |\kappa^-\rangle_{rB}|\lambda^+\rangle_{sB'}|\eta'^{(23)}\rangle_{RS} + |\kappa^-\rangle_{rB}|\lambda^-\rangle_{sB'}|\eta'^{(24)}\rangle_{RS} \\
&+ |\lambda^+\rangle_{rB}|\kappa^+\rangle_{sB'}|\eta'^{(25)}\rangle_{RS} + |\lambda^+\rangle_{rB}|\kappa^-\rangle_{sB'}|\eta'^{(26)}\rangle_{RS} + |\lambda^-\rangle_{rB}|\kappa^+\rangle_{sB'}|\eta'^{(27)}\rangle_{RS} + |\lambda^-\rangle_{rB}|\kappa^-\rangle_{sB'}|\eta'^{(28)}\rangle_{RS} \\
&+ |\lambda^+\rangle_{rB}|\lambda^+\rangle_{sB'}|\eta'^{(29)}\rangle_{RS} + |\lambda^+\rangle_{rB}|\lambda^-\rangle_{sB'}|\eta'^{(30)}\rangle_{RS} + |\lambda^-\rangle_{rB}|\lambda^+\rangle_{sB'}|\eta'^{(31)}\rangle_{RS} + |\lambda^-\rangle_{rB}|\lambda^-\rangle_{sB'}|\eta'^{(32)}\rangle_{RS}\Big\}\otimes \\
&\Big\{|\kappa^+\rangle_{tC}|\kappa^+\rangle_{uC'}|\eta'^{(33)}\rangle_{TU} + |\kappa^+\rangle_{tC}|\kappa^-\rangle_{uC'}|\eta'^{(34)}\rangle_{TU} + |\kappa^-\rangle_{tC}|\kappa^+\rangle_{uC'}|\eta'^{(35)}\rangle_{TU} + |\kappa^-\rangle_{tC}|\kappa^-\rangle_{uC'}|\eta'^{(36)}\rangle_{TU} \\
&+ |\kappa^+\rangle_{tC}|\lambda^+\rangle_{uC'}|\eta'^{(37)}\rangle_{TU} + |\kappa^+\rangle_{tC}|\lambda^-\rangle_{uC'}|\eta'^{(38)}\rangle_{TU} + |\kappa^-\rangle_{tC}|\lambda^+\rangle_{uC'}|\eta'^{(39)}\rangle_{TU} + |\kappa^-\rangle_{tC}|\lambda^-\rangle_{uC'}|\eta'^{(40)}\rangle_{TU} \\
&+ |\lambda^+\rangle_{tC}|\kappa^+\rangle_{uC'}|\eta'^{(41)}\rangle_{TU} + |\lambda^+\rangle_{tC}|\kappa^-\rangle_{uC'}|\eta'^{(42)}\rangle_{TU} + |\lambda^-\rangle_{tC}|\kappa^+\rangle_{uC'}|\eta'^{(43)}\rangle_{TU} + |\lambda^-\rangle_{tC}|\kappa^-\rangle_{uC'}|\eta'^{(44)}\rangle_{TU} \\
&+ |\lambda^+\rangle_{tC}|\lambda^+\rangle_{uC'}|\eta'^{(45)}\rangle_{TU} + |\lambda^+\rangle_{tC}|\lambda^-\rangle_{uC'}|\eta'^{(46)}\rangle_{TU} + |\lambda^-\rangle_{tC}|\lambda^+\rangle_{uC'}|\eta'^{(47)}\rangle_{TU} + |\lambda^-\rangle_{tC}|\lambda^-\rangle_{uC'}|\eta'^{(48)}\rangle_{TU}\Big\}\otimes
\end{aligned}
$$

$$\{|\kappa^+\rangle_{vD}|\kappa^+\rangle_{wD'}|\eta'^{(49)}\rangle_{VW} + |\kappa^+\rangle_{vD}|\kappa^-\rangle_{wD'}|\eta'^{(50)}\rangle_{VW} + |\kappa^-\rangle_{vD}|\kappa^+\rangle_{wD'}|\eta'^{(51)}\rangle_{VW} + |\kappa^-\rangle_{vD}|\kappa^-\rangle_{wD'}|\eta'^{(52)}\rangle_{VW}$$
$$+ |\kappa^+\rangle_{vD}|\lambda^+\rangle_{wD'}|\eta'^{(53)}\rangle_{VW} + |\kappa^+\rangle_{vD}|\lambda^-\rangle_{wD'}|\eta'^{(54)}\rangle_{VW} + |\kappa^-\rangle_{vD}|\lambda^+\rangle_{wD'}|\eta'^{(55)}\rangle_{VW} + |\kappa^-\rangle_{vD}|\lambda^-\rangle_{wD'}|\eta'^{(56)}\rangle_{VW}$$
$$+ |\lambda^+\rangle_{vD}|\kappa^+\rangle_{wD'}|\eta'^{(57)}\rangle_{VW} + |\lambda^+\rangle_{vD}|\kappa^-\rangle_{wD'}|\eta'^{(58)}\rangle_{VW} + |\lambda^-\rangle_{vD}|\kappa^+\rangle_{wD'}|\eta'^{(59)}\rangle_{VW} + |\lambda^-\rangle_{vD}|\kappa^-\rangle_{wD'}|\eta'^{(60)}\rangle_{VW}$$
$$+ |\lambda^+\rangle_{vD}|\lambda^+\rangle_{wD'}|\eta'^{(61)}\rangle_{VW} + |\lambda^+\rangle_{vD}|\lambda^-\rangle_{wD'}|\eta'^{(62)}\rangle_{VW} + |\lambda^-\rangle_{vD}|\lambda^+\rangle_{wD'}|\eta'^{(63)}\rangle_{VW} + |\lambda^-\rangle_{vD}|\lambda^-\rangle_{wD'}|\eta'^{(64)}\rangle_{VW}\}|1\rangle_E$$

(A.1)

where states $|\eta^{(i)}\rangle, i = 1 - 64$ are defined to be

$$|\eta^{(1)}\rangle = p_{00}|00\rangle + p_{01}|01\rangle + p_{10}|10\rangle + p_{11}|11\rangle, |\eta^{(2)}\rangle = p_{00}|00\rangle - p_{01}|01\rangle + p_{10}|10\rangle - p_{11}|11\rangle$$
$$|\eta^{(3)}\rangle = p_{00}|00\rangle + p_{01}|01\rangle - p_{10}|10\rangle - p_{11}|11\rangle, |\eta^{(4)}\rangle = p_{00}|00\rangle - p_{01}|01\rangle - p_{10}|10\rangle + p_{11}|11\rangle$$
$$|\eta^{(5)}\rangle = p_{00}|01\rangle + p_{01}|00\rangle + p_{10}|11\rangle + p_{11}|10\rangle, |\eta^{(6)}\rangle = p_{00}|01\rangle - p_{01}|00\rangle + p_{10}|11\rangle - p_{11}|10\rangle$$
$$|\eta^{(7)}\rangle = p_{00}|01\rangle + p_{01}|00\rangle - p_{10}|11\rangle - p_{11}|10\rangle, |\eta^{(8)}\rangle = p_{00}|01\rangle - p_{01}|00\rangle - p_{10}|11\rangle + p_{11}|10\rangle$$
$$|\eta^{(9)}\rangle = p_{00}|10\rangle + p_{01}|11\rangle + p_{10}|00\rangle + p_{11}|01\rangle, |\eta^{(10)}\rangle = p_{00}|10\rangle - p_{01}|11\rangle + p_{10}|00\rangle - p_{11}|01\rangle$$
$$|\eta^{(11)}\rangle = p_{00}|10\rangle + p_{01}|11\rangle - p_{10}|00\rangle - p_{11}|01\rangle, |\eta^{(12)}\rangle = p_{00}|10\rangle - p_{01}|11\rangle - p_{10}|00\rangle + p_{11}|01\rangle$$
$$|\eta^{(13)}\rangle = p_{00}|11\rangle + p_{01}|10\rangle + p_{10}|01\rangle + p_{11}|00\rangle, |\eta^{(14)}\rangle = p_{00}|11\rangle - p_{01}|10\rangle + p_{10}|01\rangle - p_{11}|00\rangle$$
$$|\eta^{(15)}\rangle = p_{00}|11\rangle + p_{01}|10\rangle - p_{10}|01\rangle - p_{11}|00\rangle, |\eta^{(16)}\rangle = p_{00}|11\rangle - p_{01}|10\rangle - p_{10}|01\rangle + p_{11}|00\rangle$$

$$|\eta^{(17)}\rangle = r_{00}|00\rangle + r_{01}|01\rangle + r_{10}|10\rangle + r_{11}|11\rangle, |\eta^{(18)}\rangle = r_{00}|00\rangle - r_{01}|01\rangle + r_{10}|10\rangle - r_{11}|11\rangle$$
$$|\eta^{(19)}\rangle = r_{00}|00\rangle + r_{01}|01\rangle - r_{10}|10\rangle - r_{11}|11\rangle, |\eta^{(20)}\rangle = r_{00}|00\rangle - r_{01}|01\rangle - r_{10}|10\rangle + r_{11}|11\rangle$$
$$|\eta^{(21)}\rangle = r_{00}|01\rangle + r_{01}|00\rangle + r_{10}|11\rangle + r_{11}|10\rangle, |\eta^{(22)}\rangle = r_{00}|01\rangle - r_{01}|00\rangle + r_{10}|11\rangle - r_{11}|10\rangle$$
$$|\eta^{(23)}\rangle = r_{00}|01\rangle + r_{01}|00\rangle - r_{10}|11\rangle - r_{11}|10\rangle, |\eta^{(24)}\rangle = r_{00}|01\rangle - r_{01}|00\rangle - r_{10}|11\rangle + r_{11}|10\rangle$$
$$|\eta^{(25)}\rangle = r_{00}|10\rangle + r_{01}|11\rangle + r_{10}|00\rangle + r_{11}|01\rangle, |\eta^{(26)}\rangle = r_{00}|10\rangle - r_{01}|11\rangle + r_{10}|00\rangle - r_{11}|01\rangle$$
$$|\eta^{(27)}\rangle = r_{00}|10\rangle + r_{01}|11\rangle - r_{10}|00\rangle - r_{11}|01\rangle, |\eta^{(28)}\rangle = r_{00}|10\rangle - r_{01}|11\rangle - r_{10}|00\rangle + r_{11}|01\rangle$$
$$|\eta^{(29)}\rangle = r_{00}|11\rangle + r_{01}|10\rangle + r_{10}|01\rangle + r_{11}|00\rangle, |\eta^{(30)}\rangle = r_{00}|11\rangle - r_{01}|10\rangle + r_{10}|01\rangle - r_{11}|00\rangle$$
$$|\eta^{(31)}\rangle = r_{00}|11\rangle + r_{01}|10\rangle - r_{10}|01\rangle - r_{11}|00\rangle, |\eta^{(32)}\rangle = r_{00}|11\rangle - r_{01}|10\rangle - r_{10}|01\rangle + r_{11}|00\rangle$$

$$|\eta^{(33)}\rangle = t_{00}|00\rangle + t_{01}|01\rangle + t_{10}|10\rangle + t_{11}|11\rangle, |\eta^{(34)}\rangle = t_{00}|00\rangle - t_{01}|01\rangle + t_{10}|10\rangle - t_{11}|11\rangle$$
$$|\eta^{(35)}\rangle = t_{00}|00\rangle + t_{01}|01\rangle - t_{10}|10\rangle - t_{11}|11\rangle, |\eta^{(36)}\rangle = t_{00}|00\rangle - t_{01}|01\rangle - t_{10}|10\rangle + t_{11}|11\rangle$$
$$|\eta^{(37)}\rangle = t_{00}|01\rangle + t_{01}|00\rangle + t_{10}|11\rangle + t_{11}|10\rangle, |\eta^{(38)}\rangle = t_{00}|01\rangle - t_{01}|00\rangle + t_{10}|11\rangle - t_{11}|10\rangle$$
$$|\eta^{(39)}\rangle = t_{00}|01\rangle + t_{01}|00\rangle - t_{10}|11\rangle - t_{11}|10\rangle, |\eta^{(40)}\rangle = t_{00}|01\rangle - t_{01}|00\rangle - t_{10}|11\rangle + t_{11}|10\rangle$$
$$|\eta^{(41)}\rangle = t_{00}|10\rangle + t_{01}|11\rangle + t_{10}|00\rangle + t_{11}|01\rangle, |\eta^{(42)}\rangle = t_{00}|10\rangle - t_{01}|11\rangle + t_{10}|00\rangle - t_{11}|01\rangle$$
$$|\eta^{(43)}\rangle = t_{00}|10\rangle + t_{01}|11\rangle - t_{10}|00\rangle - t_{11}|01\rangle, |\eta^{(44)}\rangle = t_{00}|10\rangle - t_{01}|11\rangle - t_{10}|00\rangle + t_{11}|01\rangle$$
$$|\eta^{(45)}\rangle = t_{00}|11\rangle + t_{01}|10\rangle + t_{10}|01\rangle + t_{11}|00\rangle, |\eta^{(46)}\rangle = t_{00}|11\rangle - t_{01}|10\rangle + t_{10}|01\rangle - t_{11}|00\rangle$$
$$|\eta^{(47)}\rangle = t_{00}|11\rangle + t_{01}|10\rangle - t_{10}|01\rangle - t_{11}|00\rangle, |\eta^{(48)}\rangle = t_{00}|11\rangle - t_{01}|10\rangle - t_{10}|01\rangle + t_{11}|00\rangle$$

$$|\eta^{(49)}\rangle = v_{00}|00\rangle + v_{01}|01\rangle + v_{10}|10\rangle + v_{11}|11\rangle, |\eta^{(50)}\rangle = v_{00}|00\rangle - v_{01}|01\rangle + v_{10}|10\rangle - v_{11}|11\rangle$$
$$|\eta^{(51)}\rangle = v_{00}|00\rangle + v_{01}|01\rangle - v_{10}|10\rangle - v_{11}|11\rangle, |\eta^{(52)}\rangle = v_{00}|00\rangle - v_{01}|01\rangle - v_{10}|10\rangle + v_{11}|11\rangle$$
$$|\eta^{(53)}\rangle = v_{00}|01\rangle + v_{01}|00\rangle + v_{10}|11\rangle + v_{11}|10\rangle, |\eta^{(54)}\rangle = v_{00}|01\rangle - v_{01}|00\rangle + v_{10}|11\rangle - v_{11}|10\rangle$$
$$|\eta^{(55)}\rangle = v_{00}|01\rangle + v_{01}|00\rangle - v_{10}|11\rangle - v_{11}|10\rangle, |\eta^{(56)}\rangle = v_{00}|01\rangle - v_{01}|00\rangle - v_{10}|11\rangle + v_{11}|10\rangle$$
$$|\eta^{(57)}\rangle = v_{00}|10\rangle + v_{01}|11\rangle + v_{10}|00\rangle + v_{11}|01\rangle, |\eta^{(58)}\rangle = v_{00}|10\rangle - v_{01}|11\rangle + v_{10}|00\rangle - v_{11}|01\rangle$$
$$|\eta^{(59)}\rangle = v_{00}|10\rangle + v_{01}|11\rangle - v_{10}|00\rangle - v_{11}|01\rangle, |\eta^{(60)}\rangle = v_{00}|10\rangle - v_{01}|11\rangle - v_{10}|00\rangle + v_{11}|01\rangle$$
$$|\eta^{(61)}\rangle = v_{00}|11\rangle + v_{01}|10\rangle + v_{10}|01\rangle + v_{11}|00\rangle, |\eta^{(62)}\rangle = v_{00}|11\rangle - v_{01}|10\rangle + v_{10}|01\rangle - v_{11}|00\rangle$$
$$|\eta^{(63)}\rangle = v_{00}|11\rangle + v_{01}|10\rangle - v_{10}|01\rangle - v_{11}|00\rangle, |\eta^{(64)}\rangle = v_{00}|11\rangle - v_{01}|10\rangle - v_{10}|01\rangle + v_{11}|00\rangle$$

and, similarly, $\left|\eta'^{(i)}\right\rangle i = 1-64$ are defined to be

$\left|\eta'^{(2)}\right\rangle = p_{00}|11\rangle - p_{01}|10\rangle - p_{10}|01\rangle + p_{11}|00\rangle, \left|\eta'^{(2)}\right\rangle = p_{00}|11\rangle + p_{01}|10\rangle - p_{10}|01\rangle - p_{11}|00\rangle$
$\left|\eta'^{(3)}\right\rangle = p_{00}|11\rangle - p_{01}|10\rangle + p_{10}|01\rangle - p_{11}|00\rangle, \left|\eta'^{(4)}\right\rangle = p_{00}|11\rangle + p_{01}|10\rangle + p_{10}|01\rangle + p_{11}|00\rangle$
$\left|\eta'^{(5)}\right\rangle = -p_{00}|10\rangle + p_{01}|11\rangle + p_{10}|00\rangle - p_{11}|01\rangle, \left|\eta'^{(6)}\right\rangle = -p_{00}|10\rangle - p_{01}|11\rangle + p_{10}|00\rangle + p_{11}|01\rangle$
$\left|\eta'^{(7)}\right\rangle = -p_{00}|10\rangle + p_{01}|11\rangle - p_{10}|00\rangle + p_{11}|01\rangle, \left|\eta'^{(8)}\right\rangle = -p_{00}|10\rangle - p_{01}|11\rangle - p_{10}|00\rangle - p_{11}|01\rangle$
$\left|\eta'^{(9)}\right\rangle = -p_{00}|01\rangle + p_{01}|00\rangle + p_{10}|11\rangle - p_{11}|10\rangle, \left|\eta'^{(10)}\right\rangle = -p_{00}|01\rangle - p_{01}|00\rangle + p_{10}|11\rangle + p_{11}|10\rangle$ ,
$\left|\eta'^{(11)}\right\rangle = -p_{00}|01\rangle + p_{01}|00\rangle - p_{10}|11\rangle + p_{11}|10\rangle, \left|\eta'^{(12)}\right\rangle = -p_{00}|01\rangle - p_{01}|00\rangle - p_{10}|11\rangle - p_{11}|10\rangle$
$\left|\eta'^{(13)}\right\rangle = p_{00}|00\rangle - p_{01}|01\rangle - p_{10}|10\rangle + p_{11}|11\rangle, \left|\eta'^{(14)}\right\rangle = p_{00}|00\rangle + p_{01}|01\rangle - p_{10}|10\rangle - p_{11}|11\rangle$
$\left|\eta'^{(15)}\right\rangle = p_{00}|00\rangle - p_{01}|01\rangle + p_{10}|10\rangle - p_{11}|11\rangle, \left|\eta'^{(16)}\right\rangle = p_{00}|00\rangle + p_{01}|01\rangle + p_{10}|10\rangle + p_{11}|11\rangle$

$\left|\eta'^{(17)}\right\rangle = r_{00}|11\rangle - r_{01}|10\rangle - r_{10}|01\rangle + r_{11}|00\rangle, \left|\eta'^{(18)}\right\rangle = r_{00}|11\rangle + r_{01}|10\rangle - r_{10}|01\rangle - r_{11}|00\rangle$
$\left|\eta'^{(19)}\right\rangle = r_{00}|11\rangle - r_{01}|10\rangle + r_{10}|01\rangle - r_{11}|00\rangle, \left|\eta'^{(20)}\right\rangle = r_{00}|11\rangle + r_{01}|10\rangle + r_{10}|01\rangle + r_{11}|00\rangle$
$\left|\eta'^{(21)}\right\rangle = -r_{00}|10\rangle + r_{01}|11\rangle + r_{10}|00\rangle - r_{11}|01\rangle, \left|\eta'^{(22)}\right\rangle = -r_{00}|10\rangle - r_{01}|11\rangle + r_{10}|00\rangle + r_{11}|01\rangle$
$\left|\eta'^{(23)}\right\rangle = -r_{00}|10\rangle + r_{01}|11\rangle - r_{10}|00\rangle + r_{11}|01\rangle, \left|\eta'^{(24)}\right\rangle = -r_{00}|10\rangle - r_{01}|11\rangle - r_{10}|00\rangle - r_{11}|01\rangle$
$\left|\eta'^{(25)}\right\rangle = -r_{00}|01\rangle + r_{01}|00\rangle + r_{10}|11\rangle - r_{11}|10\rangle, \left|\eta'^{(26)}\right\rangle = -r_{00}|01\rangle - r_{01}|00\rangle + r_{10}|11\rangle + r_{11}|10\rangle$ ,
$\left|\eta'^{(27)}\right\rangle = -r_{00}|01\rangle + r_{01}|00\rangle - r_{10}|11\rangle + r_{11}|10\rangle, \left|\eta'^{(28)}\right\rangle = -r_{00}|01\rangle - r_{01}|00\rangle - r_{10}|11\rangle - r_{11}|10\rangle$
$\left|\eta'^{(29)}\right\rangle = r_{00}|00\rangle - r_{01}|01\rangle - r_{10}|10\rangle + r_{11}|11\rangle, \left|\eta'^{(30)}\right\rangle = r_{00}|00\rangle + r_{01}|01\rangle - r_{10}|10\rangle - r_{11}|11\rangle$
$\left|\eta'^{(31)}\right\rangle = r_{00}|00\rangle - r_{01}|01\rangle + r_{10}|10\rangle - r_{11}|11\rangle, \left|\eta'^{(32)}\right\rangle = r_{00}|00\rangle + r_{01}|01\rangle + r_{10}|10\rangle + r_{11}|11\rangle$

$\left|\eta'^{(33)}\right\rangle = t_{00}|11\rangle - t_{01}|10\rangle - t_{10}|01\rangle + t_{11}|00\rangle, \left|\eta'^{(34)}\right\rangle = t_{00}|11\rangle + t_{01}|10\rangle - t_{10}|01\rangle - t_{11}|00\rangle$
$\left|\eta'^{(35)}\right\rangle = t_{00}|11\rangle - t_{01}|10\rangle + t_{10}|01\rangle - t_{11}|00\rangle, \left|\eta'^{(36)}\right\rangle = t_{00}|11\rangle + t_{01}|10\rangle + t_{10}|01\rangle + t_{11}|00\rangle$
$\left|\eta'^{(37)}\right\rangle = -t_{00}|10\rangle + t_{01}|11\rangle + t_{10}|00\rangle - t_{11}|01\rangle, \left|\eta'^{(38)}\right\rangle = -t_{00}|10\rangle - t_{01}|11\rangle + t_{10}|00\rangle + t_{11}|01\rangle$
$\left|\eta'^{(39)}\right\rangle = -t_{00}|10\rangle + t_{01}|11\rangle - t_{10}|00\rangle + t_{11}|01\rangle, \left|\eta'^{(40)}\right\rangle = -t_{00}|10\rangle - t_{01}|11\rangle - t_{10}|00\rangle - t_{11}|01\rangle$
$\left|\eta'^{(41)}\right\rangle = -t_{00}|01\rangle + t_{01}|00\rangle + t_{10}|11\rangle - t_{11}|10\rangle, \left|\eta'^{(42)}\right\rangle = -t_{00}|01\rangle - t_{01}|00\rangle + t_{10}|11\rangle + t_{11}|10\rangle$ ,
$\left|\eta'^{(43)}\right\rangle = -t_{00}|01\rangle + t_{01}|00\rangle - t_{10}|11\rangle + t_{11}|10\rangle, \left|\eta'^{(44)}\right\rangle = -t_{00}|01\rangle - t_{01}|00\rangle - t_{10}|11\rangle - t_{11}|10\rangle$
$\left|\eta'^{(45)}\right\rangle = t_{00}|00\rangle - t_{01}|01\rangle - t_{10}|10\rangle + t_{11}|11\rangle, \left|\eta'^{(46)}\right\rangle = t_{00}|00\rangle + t_{01}|01\rangle - t_{10}|10\rangle - t_{11}|11\rangle$
$\left|\eta'^{(47)}\right\rangle = t_{00}|00\rangle - t_{01}|01\rangle + t_{10}|10\rangle - t_{11}|11\rangle, \left|\eta'^{(48)}\right\rangle = t_{00}|00\rangle + t_{01}|01\rangle + t_{10}|10\rangle + t_{11}|11\rangle$

$\left|\eta'^{(49)}\right\rangle = v_{00}|11\rangle - v_{01}|10\rangle - v_{10}|01\rangle + v_{11}|00\rangle, \left|\eta'^{(50)}\right\rangle = v_{00}|11\rangle + v_{01}|10\rangle - v_{10}|01\rangle - v_{11}|00\rangle$
$\left|\eta'^{(51)}\right\rangle = v_{00}|11\rangle - v_{01}|10\rangle + v_{10}|01\rangle - v_{11}|00\rangle, \left|\eta'^{(52)}\right\rangle = v_{00}|11\rangle + v_{01}|10\rangle + v_{10}|01\rangle + v_{11}|00\rangle$
$\left|\eta'^{(53)}\right\rangle = -v_{00}|10\rangle + v_{01}|11\rangle + v_{10}|00\rangle - v_{11}|01\rangle, \left|\eta'^{(54)}\right\rangle = -v_{00}|10\rangle - v_{01}|11\rangle + v_{10}|00\rangle + v_{11}|01\rangle$
$\left|\eta'^{(55)}\right\rangle = -v_{00}|10\rangle + v_{01}|11\rangle - v_{10}|00\rangle + v_{11}|01\rangle, \left|\eta'^{(56)}\right\rangle = -v_{00}|10\rangle - v_{01}|11\rangle - v_{10}|00\rangle - v_{11}|01\rangle$
$\left|\eta'^{(57)}\right\rangle = -v_{00}|01\rangle + v_{01}|00\rangle + v_{10}|11\rangle - v_{11}|10\rangle, \left|\eta'^{(58)}\right\rangle = -v_{00}|01\rangle - v_{01}|00\rangle + v_{10}|11\rangle + v_{11}|10\rangle$
$\left|\eta'^{(59)}\right\rangle = -v_{00}|01\rangle + v_{01}|00\rangle - v_{10}|11\rangle + v_{11}|10\rangle, \left|\eta'^{(60)}\right\rangle = -v_{00}|01\rangle - v_{01}|00\rangle - v_{10}|11\rangle - v_{11}|10\rangle$
$\left|\eta'^{(61)}\right\rangle = v_{00}|00\rangle - v_{01}|01\rangle - v_{10}|10\rangle + v_{11}|11\rangle, \left|\eta'^{(62)}\right\rangle = v_{00}|00\rangle + v_{01}|01\rangle - v_{10}|10\rangle - v_{11}|11\rangle$
$\left|\eta'^{(63)}\right\rangle = v_{00}|00\rangle - v_{01}|01\rangle + v_{10}|10\rangle - v_{11}|11\rangle, \left|\eta'^{(64)}\right\rangle = v_{00}|00\rangle + v_{01}|01\rangle + v_{10}|10\rangle + v_{11}|11\rangle$

## Appendix.B

**Table 1:** Unitary operations $U_{PQ}^{(g,h)}$ & $U_{RS}^{(i,j)}$ performed by Fancy[1] and Fancy[2] (see the text below Eq.7).

| Alice's Measurement Result $(g,h)$ | Elle's Measurement Result $z$ | Fancy[1]'s operation $U_{PQ}^{(g,h)}$ | Bob's Measurement Result $(i,j)$ | Elle's Measurement Result $z$ | Fancy[2]'s operation $U_{RS}^{(i,j)}$ |
|---|---|---|---|---|---|
| (0,0) | 0 | $I \otimes I$ | (0,0) | 0 | $I \otimes I$ |
| (0,1) | 0 | $I \otimes \sigma_z$ | (0,1) | 0 | $I \otimes \sigma_z$ |
| (0,2) | 0 | $I \otimes \sigma_x$ | (0,2) | 0 | $I \otimes \sigma_x$ |
| (0,3) | 0 | $I \otimes \sigma_x \sigma_z$ | (0,3) | 0 | $I \otimes \sigma_x \sigma_z$ |
| (1,0) | 0 | $\sigma_z \otimes I$ | (1,0) | 0 | $\sigma_z \otimes I$ |
| (1,1) | 0 | $\sigma_z \otimes \sigma_z$ | (1,1) | 0 | $\sigma_z \otimes \sigma_z$ |
| (1,2) | 0 | $\sigma_z \otimes \sigma_x$ | (1,2) | 0 | $\sigma_z \otimes \sigma_x$ |
| (1,3) | 0 | $\sigma_z \otimes \sigma_x \sigma_z$ | (1,3) | 0 | $\sigma_z \otimes \sigma_x \sigma_z$ |
| (2,0) | 0 | $\sigma_x \otimes I$ | (2,0) | 0 | $\sigma_x \otimes I$ |
| (2,1) | 0 | $\sigma_x \otimes \sigma_z$ | (2,1) | 0 | $\sigma_x \otimes \sigma_z$ |
| (2,2) | 0 | $\sigma_x \otimes \sigma_x$ | (2,2) | 0 | $\sigma_x \otimes \sigma_x$ |
| (2,3) | 0 | $\sigma_x \otimes \sigma_x \sigma_z$ | (2,3) | 0 | $\sigma_x \otimes \sigma_x \sigma_z$ |
| (3,0) | 0 | $\sigma_x \sigma_z \otimes I$ | (3,0) | 0 | $\sigma_x \sigma_z \otimes I$ |
| (3,1) | 0 | $\sigma_x \sigma_z \otimes \sigma_z$ | (3,1) | 0 | $\sigma_x \sigma_z \otimes \sigma_z$ |
| (3,2) | 0 | $\sigma_x \sigma_z \otimes \sigma_x$ | (3,2) | 0 | $\sigma_x \sigma_z \otimes \sigma_x$ |
| (3,3) | 0 | $\sigma_x \sigma_z \otimes \sigma_x \sigma_z$ | (3,3) | 0 | $\sigma_x \sigma_z \otimes \sigma_x \sigma_z$ |
| (0,0) | 1 | $\sigma_x \sigma_z \otimes \sigma_x \sigma_z$ | (0,0) | 1 | $\sigma_x \sigma_z \otimes \sigma_x \sigma_z$ |
| (0,1) | 1 | $\sigma_x \sigma_z \otimes \sigma_x$ | (0,1) | 1 | $\sigma_x \sigma_z \otimes \sigma_x$ |
| (0,2) | 1 | $\sigma_x \sigma_z \otimes \sigma_z$ | (0,2) | 1 | $\sigma_x \sigma_z \otimes \sigma_z$ |
| (0,3) | 1 | $\sigma_x \sigma_z \otimes I$ | (0,3) | 1 | $\sigma_x \sigma_z \otimes I$ |
| (1,0) | 1 | $\sigma_x \otimes \sigma_x \sigma_z$ | (1,0) | 1 | $\sigma_x \otimes \sigma_x \sigma_z$ |
| (1,1) | 1 | $\sigma_x \otimes \sigma_x$ | (1,1) | 1 | $\sigma_x \otimes \sigma_x$ |
| (1,2) | 1 | $e^{i\pi}(\sigma_x \otimes \sigma_z)$ | (1,2) | 1 | $e^{i\pi}(\sigma_x \otimes \sigma_z)$ |
| (1,3) | 1 | $e^{i\pi}(\sigma_x \otimes I)$ | (1,3) | 1 | $e^{i\pi}(\sigma_x \otimes I)$ |
| (2, 0) | 1 | $\sigma_z \otimes \sigma_x\sigma_z$ | (2, 0) | 1 | $\sigma_z \otimes \sigma_x\sigma_z$ |
| (2, 1) | 1 | $e^{i\pi}(\sigma_z \otimes \sigma_x)$ | (2, 1) | 1 | $e^{i\pi}(\sigma_z \otimes \sigma_x)$ |
| (2, 2) | 1 | $\sigma_z \otimes \sigma_z$ | (2, 2) | 1 | $\sigma_z \otimes \sigma_z$ |
| (2, 3) | 1 | $\sigma_z \otimes I$ | (2, 3) | 1 | $\sigma_z \otimes I$ |
| (3, 0) | 1 | $I \otimes \sigma_x\sigma_z$ | (3, 0) | 1 | $I \otimes \sigma_x\sigma_z$ |
| (3, 1) | 1 | $e^{i\pi}(I \otimes \sigma_x)$ | (3, 1) | 1 | $e^{i\pi}(I \otimes \sigma_x)$ |
| (3, 2) | 1 | $I \otimes \sigma_z$ | (3, 2) | 1 | $I \otimes \sigma_z$ |
| (3, 3) | 1 | $I \otimes I$ | (3, 3) | 1 | $I \otimes I$ |

**Table 2**: Unitary operations $U_{TU}^{(k,l)}$ & $U_{VW}^{(m,n)}$ performed by Fancy[3] and Fancy[4] (see the text below Eq.7).

| Charlie`s Measurement Result $(k, l)$ | Elle`s Measurement Result $z$ | Fancy[3]`s operation $U_{TU}^{(k,l)}$ | David`s Measurement Result $(m, n)$ | Elle`s Measurement Result $z$ | Fancy[4]`s operation $U_{VW}^{(m,n)}$ |
|---|---|---|---|---|---|
| (0, 0) | 0 | $I \otimes I$ | (0, 0) | 0 | $I \otimes I$ |
| (0, 1) | 0 | $I \otimes \sigma_z$ | (0, 1) | 0 | $I \otimes \sigma_z$ |
| (0, 2) | 0 | $I \otimes \sigma_x$ | (0, 2) | 0 | $I \otimes \sigma_x$ |
| (0, 3) | 0 | $I \otimes \sigma_x\sigma_z$ | (0, 3) | 0 | $I \otimes \sigma_x\sigma_z$ |
| (1, 0) | 0 | $\sigma_z \otimes I$ | (1, 0) | 0 | $\sigma_z \otimes I$ |
| (1, 1) | 0 | $\sigma_z \otimes \sigma_z$ | (1, 1) | 0 | $\sigma_z \otimes \sigma_z$ |
| (1, 2) | 0 | $\sigma_z \otimes \sigma_x$ | (1, 2) | 0 | $\sigma_z \otimes \sigma_x$ |
| (1, 3) | 0 | $\sigma_z \otimes \sigma_x\sigma_z$ | (1, 3) | 0 | $\sigma_z \otimes \sigma_x\sigma_z$ |
| (2, 0) | 0 | $\sigma_x \otimes I$ | (2, 0) | 0 | $\sigma_x \otimes I$ |
| (2, 1) | 0 | $\sigma_x \otimes \sigma_z$ | (2, 1) | 0 | $\sigma_x \otimes \sigma_z$ |
| (2, 2) | 0 | $\sigma_x \otimes \sigma_x$ | (2, 2) | 0 | $\sigma_x \otimes \sigma_x$ |
| (2, 3) | 0 | $\sigma_x \otimes \sigma_x\sigma_z$ | (2, 3) | 0 | $\sigma_x \otimes \sigma_x\sigma_z$ |
| (3, 0) | 0 | $\sigma_x\sigma_z \otimes I$ | (3, 0) | 0 | $\sigma_x\sigma_z \otimes I$ |
| (3, 1) | 0 | $\sigma_x\sigma_z \otimes \sigma_z$ | (3, 1) | 0 | $\sigma_x\sigma_z \otimes \sigma_z$ |
| (3, 2) | 0 | $\sigma_x\sigma_z \otimes \sigma_x$ | (3, 2) | 0 | $\sigma_x\sigma_z \otimes \sigma_x$ |
| (3, 3) | 0 | $\sigma_x\sigma_z \otimes \sigma_x\sigma_z$ | (3, 3) | 0 | $\sigma_x\sigma_z \otimes \sigma_x\sigma_z$ |
| (0, 0) | 1 | $\sigma_x\sigma_z \otimes \sigma_x\sigma_z$ | (0, 0) | 1 | $\sigma_x\sigma_z \otimes \sigma_x\sigma_z$ |
| (0, 1) | 1 | $\sigma_x\sigma_z \otimes \sigma_x$ | (0, 1) | 1 | $\sigma_x\sigma_z \otimes \sigma_x$ |
| (0, 2) | 1 | $\sigma_x\sigma_z \otimes \sigma_z$ | (0, 2) | 1 | $\sigma_x\sigma_z \otimes \sigma_z$ |
| (0, 3) | 1 | $\sigma_x\sigma_z \otimes I$ | (0, 3) | 1 | $\sigma_x\sigma_z \otimes I$ |
| (1, 0) | 1 | $\sigma_x \otimes \sigma_x\sigma_z$ | (1, 0) | 1 | $\sigma_x \otimes \sigma_x\sigma_z$ |
| (1, 1) | 1 | $\sigma_x \otimes \sigma_x$ | (1, 1) | 1 | $\sigma_x \otimes \sigma_x$ |
| (1, 2) | 1 | $e^{i\pi}(\sigma_x \otimes \sigma_z)$ | (1, 2) | 1 | $e^{i\pi}(\sigma_x \otimes \sigma_z)$ |
| (1, 3) | 1 | $e^{i\pi}(\sigma_x \otimes I)$ | (1, 3) | 1 | $e^{i\pi}(\sigma_x \otimes I)$ |
| (2, 0) | 1 | $\sigma_z \otimes \sigma_x\sigma_z$ | (2, 0) | 1 | $\sigma_z \otimes \sigma_x\sigma_z$ |
| (2, 1) | 1 | $e^{i\pi}(\sigma_z \otimes \sigma_x)$ | (2, 1) | 1 | $e^{i\pi}(\sigma_z \otimes \sigma_x)$ |
| (2, 2) | 1 | $\sigma_z \otimes \sigma_z$ | (2, 2) | 1 | $\sigma_z \otimes \sigma_z$ |
| (2, 3) | 1 | $\sigma_z \otimes I$ | (2, 3) | 1 | $\sigma_z \otimes I$ |
| (3, 0) | 1 | $I \otimes \sigma_x\sigma_z$ | (3, 0) | 1 | $I \otimes \sigma_x\sigma_z$ |
| (3, 1) | 1 | $e^{i\pi}(I \otimes \sigma_x)$ | (3, 1) | 1 | $e^{i\pi}(I \otimes \sigma_x)$ |
| (3, 2) | 1 | $I \otimes \sigma_z$ | (3, 2) | 1 | $I \otimes \sigma_z$ |
| (3, 3) | 1 | $I \otimes I$ | (3, 3) | 1 | $I \otimes I$ |